\documentclass[aps,prl,reprint,superscriptaddress,nofootinbib,amsmath,%
amssymb,draft]{revtex4-2}
\usepackage{mathrsfs}

\begin{document}

\title{Unveiling the electrodynamics of the first nonlinearly charged
rotating black hole}

\author{Eloy Ay\'on-Beato}
\email{ayon-beato-at-fis.cinvestav.mx}
\affiliation{Departamento de F\'{\i}sica,
CINVESTAV-IPN, Apartado Postal 14740, 07360 M\'exico D.F., M\'exico}

\begin{abstract}
After many years of efforts, the first nonlinearly charged rotating black
hole has been finally reported by Garc\'{\i}a-Diaz in two recent works.
This is an important result that was pending in General Relativity, since
nonlinear generalizations of the Kerr-Newman solution were not yet known.
Unfortunately, the Lagrangian supporting this configuration cannot be
expressible in terms of the standard invariants using elementary functions.
In the present work we circumvent this problem by using the formulations of
nonlinear electrodynamics in terms of mixed electromagnetic eigenvalues,
introduced by Salazar, Garc\'{\i}a-Diaz and Pleba\'nski almost four decades
ago. In doing so, we prove that the underlying theory becomes fully
determined, and hence the new found nonlinearly charged stationary
axisymmetric spacetimes correspond to exact solutions of a well-defined
self-gravitating nonlinear electrodynamics whose fundamental structural
functions are provided here.
\end{abstract}

\maketitle

In their seminal work on nonlinear electrodynamics \cite{Salazar:1987ap}
Salazar, Garc\'{\i}a-Diaz and Pleba\'nski conclude: ``\emph{\ldots We
consider a derivation of such solutions, which would generalize the
Kerr-Newman solution for the case of the nonlinear rotating charges as an
open challenging problem within the theory of exact solutions in general
relativity.}'' Almost four decades later, after tireless and diverse efforts,
not only by the Cinvestav group but by many others around the world, this
challenge has been finally overtaken by Professor Garc\'{\i}a-Diaz in two
recent groundbreaking papers \cite{Garcia-Diaz:2021bao,Diaz:2022roz}.

His main hypothesis is the alignment of the metric null tetrad along the
common eigenvectors of the electromagnetic fields. This condition was also
assumed in \cite{Salazar:1987ap} and in fact since the foundational book of
Pleba\'nski on the subject \cite{Plebanski:1970zz}. However, its principal
consequence had not been fully explored until the work of Garc\'{\i}a-Diaz
\cite{Garcia-Diaz:2021bao,Diaz:2022roz}: the complete separability of the
stationary axisymmetric electromagnetic fields. Concretely, the alignment
condition implies that given a null tetrad for the metric {\small
\begin{equation}\label{eq:NullT}
g=2 \theta^1\otimes_\text{s}\theta^2 + 2 \theta^3\otimes_\text{s}\theta^4,
\end{equation}}%
where the first pair of null one-forms are complex conjugates while the last
two are real, the electromagnetic closed two-forms embodying the Faraday and
Maxwell equations \cite{Born:1934gh} {\small
\begin{equation}\label{eq:FM}
dF=0 \qquad \text{and} \qquad d*P=0,
\end{equation}}%
respectively, are necessarily decomposed as {\small
\begin{equation}\label{eq:F*Pa}
F+i*P=(D+iB)\theta^1\!\wedge\theta^2 + (E+iH)\theta^3\!\wedge\theta^4.
\end{equation}}%
This means the null tetrad \eqref{eq:NullT} are the common eigenvectors of
the electromagnetic fields and the unique tetrad components $E$, $B$ and $D$,
$H$ determine their eigenvalues, being real alternative invariants physically
associated in the first pair to the intensity of electric field and magnetic
induction and in the second to the electric induction and intensity of
magnetic field \cite{Plebanski:1970zz}. They allow to express the standard
invariants as {\small
\begin{subequations}\label{eq:Invs}
\begin{align}
\mathscr{F}+i\mathscr{G}&\equiv
\frac14 F_{ab}F^{ab}+\frac{i}4 F_{ab}*F^{ab} =
-\frac12(E+iB)^2, \label{eq:InvsFG} \\
\mathscr{P}+i\mathscr{Q}&\equiv
\frac14 P_{ab}P^{ab}+\frac{i}4 P_{ab}*P^{ab} =
-\frac12(D+iH)^2. \label{eq:InvsPQ}
\end{align}
\end{subequations}}%
Hence, as was brilliantly unveiled in \cite{Salazar:1987ap} the description
of nonlinear electrodynamic theories is not exhausted by their determination
in terms of fundamental structural functions depending on the standard
invariants as the Lagrangian $\mathscr{L}(\mathscr{F},\mathscr{G})$ or its
Legendre transform, the ``Hamiltonian''
$\mathscr{H}(\mathscr{P},\mathscr{Q})=\frac12F_{\mu\nu}P^{\mu\nu}-\mathscr{L}$
\cite{Born:1934gh}. Instead, thanks to the aligned tetrad invariants, it is
not only possible to reformulate the theories using $\mathscr{L}(E,B)$ or
$\mathscr{H}(D,H)$ \cite{Plebanski:1970zz}, but also extra formulations are
conceivable using mixed fundamental structural functions obtained from the
subsequent Legendre transforms $\mathscr{M}^{(+)}(D,B)=BH-\mathscr{H}$ and
$\mathscr{M}^{(-)}(E,H)=DE+\mathscr{H}$ depending on the inductions and
intensities, respectively \cite{Salazar:1987ap}.

The relevance of the above discussion lies in the following. In his second
work \cite{Diaz:2022roz} Garc\'{\i}a-Diaz proved that there exists a
Lagrangian supporting the nonlinearly charged generalization of the
Kerr-Newman black hole reported in his first work \cite{Garcia-Diaz:2021bao},
which would imply this is the first analytic stationary axisymmetric solution
for nonlinear electrodynamics found in the Literature. Unfortunately, he
correctly argues that this Lagrangian is not expressible in terms of the
standard invariants using elementary functions. In this work, we show that it
is precisely in one of the mentioned mixed formulations where the related
electrodynamics becomes fully determined.

The most transparent action principle describing the full dynamics of a
self-gravitating nonlinear electrodynamics is derived from
\cite{Plebanski:1970zz,Salazar:1987ap} {\small
\begin{align}
S[g^{\mu\nu},A_{\mu},P^{\mu\nu}]={}&\int d^4x\sqrt{-g}\left[\frac1{16\pi}R\right.
\nonumber \\
&-\left.\frac1{4\pi}\left(\frac12F_{\mu\nu}P^{\mu\nu}
-\mathscr{H}(\mathscr{P},\mathscr{Q})\right)\right],\label{eq:ActionNLE}
\end{align}}%
where $R$ stands for the scalar curvature of the metric $g_{\mu\nu}$ given by
the trace of the Ricci tensor
$R_{\mu\nu}=R^\alpha_{~\mu\alpha\nu}$.\footnote{We use the custom notation of
Misner, Thorne and Wheeler book, that differs in a sign with the employed in
\cite{Salazar:1987ap}.} The Maxwell--Faraday equations \eqref{eq:FM} are
interconnectedly considered in the action principle as follows: on the one
hand, according to the Faraday equation in \eqref{eq:FM}, the field strength
is necessarily expressed in terms of a vector potential
$F_{\mu\nu}=\partial_{\mu}A_{\nu}-\partial_{\nu}A_{\mu}$, on the other hand
varying the action with respect to $A_{\mu}$ gives precisely the Maxwell
equation in \eqref{eq:FM}. Since the resulting Maxwell equations are linear
in terms of the conjugate antisymmetric tensor $P^{\mu\nu}$, the nonlinear
contents is now encoded in the variation of the action with respect to this
field, which gives rise to the constitutive or material relations
\begin{equation}\label{eq:ConstiRel}
F_{\mu\nu} = \mathscr{H}_\mathscr{P} P_{\mu\nu}
+\mathscr{H}_\mathscr{Q} *P_{\mu\nu},
\end{equation}
between both fields, where the ``Hamiltonian'' fundamental structural
function $\mathscr{H}(\mathscr{P},\mathscr{Q})$ defines the concrete
electrodynamics. For example, the well-known Maxwell linear theory is given
by $\mathscr{H}_\text{M}=\mathscr{P}$. Finally, the self-gravitating behavior
is obtained from the metric variation, giving Einstein equations
\begin{subequations}\label{eq:Einstein}
\begin{equation}
G_{\mu\nu}=8\pi T_{\mu\nu},
\end{equation}
for the energy--momentum tensor
\begin{equation}\label{eq:Tmunu}
4\pi T_{\mu\nu}=F_{\mu\alpha}P_\nu^{~\alpha}
-g_{\mu\nu}\left(\frac12F_{\alpha\beta}P^{\alpha\beta}
-\mathscr{H}\right).
\end{equation}
\end{subequations}
The advantage of this action principle, with respect to the more
straightforward involving the Lagrangian
$\mathscr{L}(\mathscr{F},\mathscr{G})$ instead of its ``Hamiltonian''
Legendre transform, is that it not only decomposes the complexity of
nonlinear electrodynamics in simpler ingredients as shown, but also provides
a schematic methodology for the searching of self-gravitating configurations:
first, solve for $P_{\mu\nu}$ the now linear Maxwell equation in
\eqref{eq:FM}, second, inserting the result in the constitutive relations
\eqref{eq:ConstiRel} of a given theory the nonlinear electromagnetic strength
$F_{\mu\nu}$ is found, and third, with both results build the energy-momentum
tensor \eqref{eq:Tmunu} in order to solve Einstein equations
\eqref{eq:Einstein} (incidentally, Garc\'{\i}a-Diaz has introduced a new
approach in his recent work \cite{Diaz:2022roz} that we will discuss later).
However, it is crucial to emphasize that there are cases where the
equivalence between both formalisms is not necessarily warranted; since the
structural functions are Legendre transforms they are equivalent only when
their conjugate relations, embodied here in the constitutive relations
\eqref{eq:ConstiRel}, are invertible \cite{Plebanski:1970zz}. We assume that
such inversion is possible, which not necessarily implies it is expressed
through elementary functions. Interesting examples of electrodynamics
properly defined in terms of the ``Hamiltonian''
$\mathscr{H}(\mathscr{P},\mathscr{Q})$, describing physically sensible
nonlinearly charged configurations, but not allowing Lagrangians
$\mathscr{L}(\mathscr{F},\mathscr{G})$ being single elementary functions of
the first pair of invariants \eqref{eq:InvsFG} are known. Particularly
outstanding cases are those related to regular black holes
\cite{Ayon-Beato:1998hmi} or more recently to Lifshitz black holes
\cite{Alvarez:2014pra}. This is precisely the problem with the nonlinearly
charged rotating black holes recently reported by Garc\'{\i}a-Diaz
\cite{Garcia-Diaz:2021bao,Diaz:2022roz}, even worse the problem extend to the
described formulation in terms of $\mathscr{H}(\mathscr{P},\mathscr{Q})$
since an elementary dependence on the second pair of invariants
\eqref{eq:InvsPQ} does not seems possible either. It is precisely in this
kind of situations where the mixed formulations of nonlinear electrodynamics
introduced in \cite{Salazar:1987ap} become relevant.

Before describing these formulations, it is pertinent to rewrite the derived
field equations in the base of the aligned null tetrad, defined by
\eqref{eq:NullT} and \eqref{eq:FM}, in addition to changing the dependence on
the invariants to $\mathscr{H}(D,H)$, see
\cite{Plebanski:1970zz,Salazar:1987ap}. First, it is straightforward to
substitute the aligned fields \eqref{eq:F*Pa} into the Maxwell-Faraday
equations \eqref{eq:FM} and use the first Cartan equations for the null
tetrad \eqref{eq:NullT}. The resulting expressions are of little use here and
can be consulted in \cite{Salazar:1987ap}. After using decomposition
\eqref{eq:F*Pa}, the constitutive relations \eqref{eq:ConstiRel} become
\begin{equation}\label{eq:ConstiRelEB(DH)}
E+iB=(-\partial_D+i\partial_H)\mathscr{H}.
\end{equation}
Regarding Einstein equations \eqref{eq:Einstein}, following
\cite{Salazar:1987ap} it is useful to break them down in their traceless part
and trace
\begin{subequations}\label{eq:EinsteinNT}
\begin{align}
S&=2(DE+BH)(\theta^1\otimes_\text{s}\theta^2-\theta^3\otimes_\text{s}\theta^4),
\label{eq:EinsteinTL}\\
R&=-4(DE-BH)-8\mathscr{H},\label{eq:EinsteinT}
\end{align}
\end{subequations}
where the tensor $S_{ab}\equiv R_{ab}-\frac14Rg_{ab}$ is the traceless part
of the Ricci tensor and the fields $E$ and $B$ are calculated from the
constitutive relations \eqref{eq:ConstiRelEB(DH)}.

We are now in a position to reformulate the equations using mixed structural
functions defined by the subsequent Legendre transforms originally introduced
in \cite{Salazar:1987ap}
\begin{equation}\label{eq:Mpm}
\mathscr{M}^{(+)}=BH-\mathscr{H}, \qquad
\mathscr{M}^{(-)}=DE+\mathscr{H}.
\end{equation}
Notice that the constitutive relations \eqref{eq:ConstiRelEB(DH)} are
equivalent to expressing the differential of $\mathscr{H}$ as
\begin{equation}\label{eq:dH}
d\mathscr{H}=-EdD+BdH,
\end{equation}
that in turn specifies the other differentials by
\begin{equation}\label{eq:dMpm}
d\mathscr{M}^{(+)}=EdD+HdB, \quad
d\mathscr{M}^{(-)}=DdE+BdH,
\end{equation}
giving the new forms of the constitutive relations for each formulation
{\small
\begin{equation}\label{eq:ConstiRelEH(DB)}
E+iH=(\partial_D+i\partial_B)\mathscr{M}^{(+)}, \quad
D+iB=(\partial_E+i\partial_H)\mathscr{M}^{(-)},
\end{equation}}%
and allowing to conclude that the functional dependence of the recently
introduced structural functions is in fact mixed, i.e.\
$\mathscr{M}^{(+)}=\mathscr{M}^{(+)}(D,B)$ is a function of the electric and
magnetic inductions and $\mathscr{M}^{(-)}=\mathscr{M}^{(-)}(E,H)$ is a
function of the electric and magnetic intensities. For example, the linear
Maxwell theory is recovered for
\begin{equation}\label{eq:Maxwell}
\mathscr{M}^{(+)}_\text{M}=\frac12(D^2+B^2), \quad
\mathscr{M}^{(-)}_\text{M}=\frac12(E^2+H^2).
\end{equation}
The Einstein equations \eqref{eq:EinsteinNT} for these
formulations are
\begin{subequations}\label{eq:EinsteinM+-}
\begin{align}
S&=2(DE+BH)(\theta^1\otimes_\text{s}\theta^2-\theta^3\otimes_\text{s}\theta^4),
\label{eq:EinsteinTLM+-}\\
R&=\pm8\mathscr{M}^{(\pm)}\mp4(DE+BH),\label{eq:EinsteinTM+-}
\end{align}
\end{subequations}
where now the derived electromagnetic fields must be calculated from the
appropriate constitutive relations \eqref{eq:ConstiRelEH(DB)} depending if we
are working with the induction or intensity formulations.

The electrodynamics we propose to support the nonlinearly charged stationary
axisymmetric configurations recently reported in
\cite{Garcia-Diaz:2021bao,Diaz:2022roz} is determined by any of the following
induction and intensity dependent structural functions
\begin{subequations}\label{eq:MpMmG}  {\small
\begin{equation}
\mathscr{M}^{(\pm)}-\mathscr{M}^{(\pm)}_\text{M}\pm\frac34\beta^2(q^2+p^2)=
\begin{cases}
f(D,B),\\
f(E,H),
\end{cases}
\end{equation}}%
where the common two-argument function is {\small
\begin{equation}\label{eq:f}
f(x,y)=\frac{\beta^2}4(q^2+p^2)
-\beta\sqrt{(q^2+p^2)\left[(x+\beta q)^2+(y-\beta p)^2\right]}.
\end{equation}}%
\end{subequations}
This theory becomes the Maxwell limit \eqref{eq:Maxwell} when the coupling
constant $\beta$ vanishes, describing the more interesting sector of the
configurations \cite{Garcia-Diaz:2021bao,Diaz:2022roz} where this limit is
clearly achieved. In order to check that they are in fact exact solutions for
this nonlinear electrodynamics, it is time to outline now the clever new
method devised by Garc\'{\i}a-Diaz to understand the stationary axisymmetric
sector of this kind of theories \cite{Diaz:2022roz}.

First, the general solution to the Maxwell-Faraday equations \eqref{eq:FM}
implies the local existence of a pair of vector potentials
\begin{equation}\label{eq:FMsol}
F=dA, \qquad *P=dA^*.
\end{equation}
Second, in a stationary axisymmetric spacetime with Killing vectors
$\partial_t$ and $\partial_\phi$, these symmetries are only required in the
gauge invariant field strengths $F$ and $*P$ and not necessarily in the gauge
fields $A$ and $A^*$. However, it is possible to argue that there exist
common gauges where both vector potentials are expressible as
\begin{equation}\label{eq:AA*StatAxis}
A=A_t dt+A_\phi d\phi, \qquad A^*=A^*_t dt+A^*_\phi d\phi,
\end{equation}
with components independent of the Killing coordinates. Third, using this
gauge to build the strengths \eqref{eq:FMsol} and changing to any preselected
stationary axisymmetric null base \eqref{eq:NullT} does not warrant to
identically satisfies the decomposition \eqref{eq:F*Pa}. The imposition of
\eqref{eq:F*Pa} generates the alignment conditions: the demand that the
components not appearing in \eqref{eq:F*Pa} vanish, constraining the
stationary axisymmetric vector potentials \eqref{eq:AA*StatAxis}. Remarkably,
these constrains are integrable for the stationary axisymmetric spacetimes
studied by Garc\'{\i}a-Diaz, and in general for the Carter-Pleba\'nski
separable metric class \cite{Carter:1968ks,Plebanski:1975xfb}. They give the
same separable ansatz for the vector potentials that is obtained by Carter
after demanding separability of charged Klein-Gordon equations, where the
potentials are minimally coupled \cite{Carter:1968ks,Carter:1973rla}; the
potential allowing the separability is necessarily determined up to a pair of
single variable functions, one for each non-Killing coordinate. Fourth, this
provides precise expressions for the nontrivial tetrad components
\eqref{eq:F*Pa} which must be further constrained for the electrodynamics of
interest by the constitutive relations \eqref{eq:ConstiRelEH(DB)}, since the
right hand sides of the one-forms \eqref{eq:dMpm} constructed with them are
not necessarily exact forms a priory. Denoting these inexact forms as
$\delta\mathscr{M}^{(\pm)}$, a necessary and sufficient condition for which
the obtained vector potentials be related to some electrodynamics, and
consequently these one-forms can be properly expressed as exact
differentials, is to demand that they are closed,
$d\delta\mathscr{M}^{(\pm)}=0$, this is what Garc\'{\i}a-Diaz dubbed the
\emph{KEY} conditions and are independent of the chosen formulation. The
single variable functions satisfying them give rise to the vector potentials
of a given electrodynamics, whose structural function can be integrated in
terms of local coordinates. It is important to emphasize that this
integration always produces an arbitrary integration constant in the
structural function, which at the level of action \eqref{eq:ActionNLE} can be
reinterpreted as a cosmological constant. Reexpressing the thus obtained
structural function in terms of the corresponding invariants is the remaining
goal which, as the Garc\'{\i}a-Diaz explicit example shows
\cite{Garcia-Diaz:2021bao,Diaz:2022roz}, is not necessarily achieved in the
$\mathscr{L}(E,B)$ or $\mathscr{H}(D,H)$ formulations. In his example,
Garc\'{\i}a-Diaz chose cubic dependent single variable functions to satisfy
the \emph{KEY} conditions and the resulting Lagrangian cannot be rewritten as
a function of the invariants using elementary functions. Fortunately, this is
not the case for the mixed formulations where the induction and intensity
dependent structural functions have the simple expressions \eqref{eq:MpMmG}.
The fifth and last step, is to solve Einstein equations
\eqref{eq:EinsteinM+-} for the involved gravitational potentials, since after
the previous steps all the components of the energy-momentum tensor
\eqref{eq:Tmunu} are written as functions of the local coordinates.

The Garc\'{\i}a-Diaz black hole obtained by following the above guidelines
in \cite{Garcia-Diaz:2021bao} is
\begin{subequations}\label{eq:GarciaSol}
\begin{align}
ds^2={}&\Sigma d\theta^2
+\frac{\sin^2\theta}{\Sigma}\left(a d t-(r^2+a^2)d\phi\right)^2\nonumber\\
&+\frac{\Sigma}{\Delta}dr^2
-\frac{\Delta}{\Sigma}\left(d t-a\sin^2\theta d\phi\right)^2,
\label{eq:GarciaMetric}\\
A={}&\frac{p\cos\theta\left(1-\beta a^2\sin^2\theta\right)}{\Sigma}
\left(a d t-(r^2+a^2)d\phi\right)\nonumber\\
&-\frac{qr\left(1-\beta(r^2+a^2)\right)}{\Sigma}\left(d t-a\sin^2\theta d\phi\right),
\label{eq:GarciaA}\\
A^*={}&\frac{q\cos\theta\left(1-\beta a^2\sin^2\theta\right)}{\Sigma}
\left(a d t-(r^2+a^2)d\phi\right)\nonumber\\
&+\frac{pr\left(1-\beta(r^2+a^2)\right)}{\Sigma}\left(d t-a\sin^2\theta d\phi\right),
\label{eq:GarciaA*}
\end{align}
where $\Sigma=r^2+a^2\cos^2\theta$ and\footnote{Our constants are related to
those of Garc\'{\i}a-Diaz by $q=q_\text{G}(1+\beta_\text{G}a^2)$,
$p=p_\text{G}(1+\beta_\text{G}a^2)$ and
$\beta=\beta_\text{G}/(1+\beta_\text{G}a^2)$.}
\begin{equation}
\Delta=r^2+a^2-2mr+(q^2+p^2)\left(1-\beta(r^2+a^2)\right)^2.
\end{equation}
\end{subequations}
For $\beta=0$ the linearly charged Kerr-Newman black hole
\cite{Newman:1965my}, in its dyonic version \cite{Demianski:1966}, is
consistently recovered. The corresponding aligned null tetrad
\eqref{eq:NullT} can be straightforwardly read off from
\eqref{eq:GarciaMetric} by recognizing the square sum in its first line as
the tensorial ``modulus'' of a complex one-form, and by factorizing the
squares difference of its second line. The aligned null tetrad
electromagnetic invariants \eqref{eq:F*Pa} are then build from the time
component of the vector potentials \eqref{eq:GarciaA} and \eqref{eq:GarciaA*}
as
\begin{subequations}\label{eq:GarciaEBDH}
\begin{align}
D&=\frac1{a\sin\theta}\partial_\theta{A^*_t}, &
B&=-\frac1{a\sin\theta}\partial_\theta{A_t}, \\
E&=-\partial_r{A_t}, & H&=-\partial_r{A^*_t}.
\end{align}
\end{subequations}
It is straightforward to check that the resulting expressions satisfy the
constitutive relations \eqref{eq:ConstiRelEH(DB)} for the structural
functions \eqref{eq:MpMmG}. Since Einstein equations \eqref{eq:EinsteinM+-}
are additionally satisfied, the configuration \eqref{eq:GarciaSol} found by
Garc\'{\i}a-Diaz in \cite{Garcia-Diaz:2021bao} is in fact a genuine
stationary axisymmetric \emph{exact} solution of the
Einstein-nonlinear-electrodynamics system for the precise theory well-defined
by the structural functions \eqref{eq:MpMmG}. It deserves all the credit as
the first solution with these properties in the whole Literature. Its
generalization reinterpreting the posible arbitrary constant of the
structural function as a cosmological constant is characterized along similar
lines and generates the second solution \cite{Diaz:2022roz}, which is
consequently supported by the same electrodynamics \eqref{eq:MpMmG}.

Regarding the physical interpretation of these authentic \emph{exact}
solutions it is enlightening to emphasize their nonlinearly charged
character. A careful evaluation gives in both solutions
\begin{equation}\label{eq:q,p}
\frac1{4\pi}\int_{S^2}*P=q, \qquad
\frac1{4\pi}\int_{S^2}F=p,
\end{equation}
where $S^2$ is any sphere (not necessarily at infinity), so that the
integration constants $q$ and $p$ are in fact the electric and magnetic
charges of the nonlinear regime, respectively. This allows to assure,
paraphrasing the famous quote of \cite{Salazar:1987ap} with which we started,
that the methods and resulting solutions recently discovered by
Garc\'{\i}a-Diaz in \cite{Garcia-Diaz:2021bao,Diaz:2022roz} \emph{generalize
the Kerr-Newman solution for the case of the nonlinear rotating charges}
opening the door to understand many more of such configurations, so that they
can be considered without a doubt as a milestone \emph{within the theory of
exact solutions in general relativity}.

\begin{acknowledgments}
I thank Professor A.~Garc\'{\i}a-Diaz for opening the doors for me to the
intricacies of nonlinear electrodynamics and for keeping me up to date on his
progress on the problem through the years. Recent discussions with
D.~Flores-Alfonso and M.~Hassa\"{\i}ne, as well as their corrections in the
present manuscript, are also highly appreciated. This work has been partially
funded by Grant No.~A1-S-11548 from Conacyt.
\end{acknowledgments}

\end{document}